\documentclass[superscriptaddress,reprint]{revtex4-2}

\usepackage{xcolor}
\usepackage{amsmath}    % Fancy math fonts
\usepackage{graphicx}   % Include figure files
\usepackage{dcolumn}    % Align table columns on decimal point
\usepackage{bm}         % Bold math
\usepackage[caption=false]{subfig}
\usepackage{hyperref}
\newcommand{\xm}{\relax\ifmmode X_{\mathrm{max}} \else
  $X_{\mathrm{max}}$\fi}
\newcommand{\mxm}{\relax\ifmmode \left<X_{\mathrm{max}}\right> \else
  $\left<X_{\mathrm{max}}\right>$\fi}
\newcommand{\sxm}{\relax\ifmmode \sigma(X_{\mathrm{max}}) \else
  $\sigma(X_{\mathrm{max}})$\fi}
\newcommand{\nm}{\relax\ifmmode N_{\mathrm{max}} \else
  $N_{\mathrm{max}}$\fi}

\begin{document}
\title{First High-speed Video Camera Observations of a Lightning Flash Associated with a Downward Terrestrial Gamma-ray Flash}
%%%%%%%%%%%%%%%%%%%%%%%%%%%%%%%%%%%%%%%
%\input{TA-author-20200522}

\author{R.U. Abbasi}
\email{rabbasi@luc.edu}
\affiliation{Department of Physics, Loyola University Chicago, Chicago, Illinois, USA}

\author{M. M. F. Saba}
\affiliation{National Institute for Space Research (INPE), Sao Jose dos Campos, Brazil}

\author{J.W.Belz}
\affiliation{Department of Physics and Astronomy , University of Utah, Sale Lake City , Utah, USA}

\author{P. R. Krehbiel}
\affiliation{Langmuir Laboratory for Atmospheric Research, New Mexico Institute of Mining and Technology, Socorro, NM, USA}

\author{W. Rison}
\affiliation{Langmuir Laboratory for Atmospheric Research, New Mexico Institute of Mining and Technology, Socorro, NM, USA}

\author{N. Kieu}
\affiliation{Department of Physics, Loyola University Chicago, Chicago, Illinois, USA}

\author{D. R. da Silva}
\affiliation{National Institute for Space Research (INPE), Sao Jose dos Campos, Brazil}

\author{Dan Rodeheffer} 
\affiliation{Langmuir Laboratory for Atmospheric Research, New Mexico Institute of Mining and Technology, Socorro, NM, USA}

\author{M. A. Stanley}
\affiliation{Langmuir Laboratory for Atmospheric Research, New Mexico Institute of Mining and Technology, Socorro, NM, USA}

\author{J. Remington}
\affiliation{Department of Physics and Astronomy , University of Utah, Sale Lake City , Utah, USA}
\affiliation{NASA Marshall Space Flight Center, Huntsville, Alabama, USA}

% ...

%\input{TA-author-20211010-aastex_copy}
%\input{TA-author-20220309}

%%%%%%%%%%%%%%%%%%%%%%%%%%%%%%%%%%%%%%%%


%\linenumbers\relax            % Comment out to disable line numbering
\begin{abstract} 

In this paper, we present the first high-speed video observation of a cloud-to-ground lightning flash and its associated downward-directed Terrestrial Gamma-ray Flash (TGF). The optical emission of the event was observed by a high-speed video camera running at 40,000 frames per second in conjunction with the Telescope Array Surface Detector, Lightning Mapping Array, interferometer, electric-field fast antenna, and the National Lightning Detection Network. The cloud-to-ground flash associated with the observed TGF was formed by a fast downward leader followed by a very intense return stroke peak current of -154 kA. The TGF occurred while the downward leader was below cloud base, and even when it was halfway in its propagation to ground. The suite of gamma-ray and lightning instruments, timing resolution, and source proximity offer us detailed information and therefore a unique look at the TGF phenomena

\end{abstract}

 %rm for apj
\maketitle % rm for apj
%%%%%%%%%%%%%%%%%%%%%%%%%%%%%%%%%%%%%%
%\input{introduction}

\section*{Introduction}

Terrestrial Gamma-ray Flashes (TGFs) are bursts of gamma-ray radiation
produced via bremsstrahlung in the Earth's atmosphere. TGFs were first
observed by the Burst and Transient Source Experiment (BATSE) on the
Compton Gamma-Ray Observatory satellite~\cite{fishman1994,BATSE1994}. The leading mechanism that produces TGFs is believed to be the Relativistic Runaway Electron Avalanche (RREA)~\cite{Gurevich61,GUREVICH1992463}.  The two main theories were developed to interpret the high-fluence TGF observations are the Relativistic Feedback  Discharge mechanism (RFD)~\cite{Dwyer2003,https://doi.org/10.1029/2011JA017160,https://doi.org/10.1002/grl.50742} and the lightning leader models also called the thermal runaway mechanism~\cite{Celestin2011}. 

In order to understand the physics behind the initiation and propagation of TGFs, satellite experiments (e.g. RHESSI~\cite{smith2005terrestrial,RHESSI2,RHESSI1}, Fermi~\cite{Fermi}, AGILE~\cite{agile}, ASIM~\cite{asimmiss2019neub, Heumesser21,Lindanger22, Engeland22}) have detected thousands of TGFs, and more recently several  downward-directed TGFs have also been observed from the ground~\cite{Tran2015,Hare2016,Enoto2017,abbasi2018gamma,Pleshinger19,belz2020observations}. 
Despite several TGF observations being reported in the literature, the mechanism responsible for producing them, and  how IC/CG discharges relate to TGFs is still not fully understood.

The sequence of the optical observations of lightning flashes in association with TGFs and in conjunction with high/low radio frequency emissions can  improve our understanding of the development stage of the lightning discharge when TGFs occur. It could also enhance our understanding of TGF initiation. The RFD mechanism, introduced by~\citep{Dwyer2003,https://doi.org/10.1002/grl.50742}, suggests that photons and positrons produce feedback that exponentially increases the number of runaway electrons. If TGFs are byproducts of \textit{relativistic feedback discharges}, they will consist of high-current electric discharges that generate radio emissions similar to lightning. This will produce light in the UV lines. ~\citep{dwyer2013properties} referred to the RFD process as ``dark lightning". The thermal runaway electron's production mechanism, on the other hand, assumes very localized regions in space (streamer heads) of a high electric field (ten times larger than the conventional breakdown field)~\cite{DwyerUman}. If TGFs are produced through \textit{thermal runaway electrons} in streamers, then this discharge should produce an optical signal before or simultaneously with the production of gamma-ray emission~\cite{Xu2015}. These scenarios together with the emission sequence are still under investigation~\cite{PhysRevD.100.083018}.

Recent observations from the Atmosphere-Space Interactions Monitor (ASIM)~\cite{asimmiss2019neub, Heumesser21,Lindanger22, Engeland22} have revealed, for the first time and, with high timing accuracy, the optical emission timing and strength associated with lightning discharges in coincidence with TGF and Transient Luminous Event (TLE) observations. The Modular Multi-spectral Imaging Array (MMIA)~\cite{mmia2019} on-board ASIM~\cite{asimmiss2019neub}, is made of two cameras and high-speed photometers at 337~nm, and 777.4~nm (both used for detecting lightning) with a 100 kHz sampling rate~\cite{asim10m,skie22}.

This work presents the first simultaneous detection of a downward TGF together with the observation of the associated cloud-to-ground lightning flash by a high-speed camera. The camera, operating at 40,000 images per second, allowed us to examine the development stage of the lightning flash during the occurrence of a TGF. The advantage of proximity to the source and the use of a suite of lightning instruments together with the high-speed camera made possible some further understanding of the characteristics of lightning processes associated with TGF production. It also allowed us to compare the optical emission observations of lightning discharges associated with downward vs. upward-moving TGFs.

\section*{Instrumentation}
\label{TA}

The observations of downward TGFs, reported in this work, were detected by the Telescope Array Surface Detector (TASD) located in the southwestern desert of Utah. The flashes that produced the TGFs were recorded simultaneously by a high-speed video camera, a Lightning Mapping Array (LMA), a broadband VHF interferometer, and Fast Antenna (FA).  Figure~S1 (provided in the supporting information) shows the layout of all of the involved detectors. Each of these detectors is described in detail in~\citep{abu2012surface,abbasi2017bursts,abbasi2018gamma,belz2020observations}, and ~\citep{saba2022}.  In this section, we will introduce each of these instruments briefly. 

The \textbf{Telescope Array Surface Detector (TASD)} is a ground-based surface detector primarily designed to observe Ultra High Energy Cosmic Rays (UHECR). With an area coverage of 700 km$^2$ the TASD is the largest  UHECR detector in the Northern Hemisphere. It comprises 507 Surface Detectors (SDs) installed at 1.4 km MSL. Each SD unit consists of upper and lower scintillator planes. Each plane is 3 m$^2$ in area and  1.2 cm in  thickness, separated by a 1 mm thick stainless steel sheet. Each plane is read out by individual photomultiplier tubes via wavelengths shifting fiber. The output  signals from the photomultipliers are digitized at a 50 MHz sampling rate.  Each SD is housed inside an RF-sealed and light-tight stainless-steel enclosure. An event trigger occurs when three adjacent SDs observe a signal greater than 3 Minimum Ionizing Particles within 8 $\mu$s ($\sim$ 150 FADC counts). When an event trigger occurs, the signals from all individually triggered SDs within~$\pm$32 $\mu$s are recorded \cite{abu2012surface,abbasi2018gamma}. The TASD detector observed approximately 25 downward terrestrial gamma-ray flashes within the past 13 years, making it the world-leading detector in downward TGF observations. Details of the detector's design, trigger, and particle energy can be found in~\cite{abu2012surface,abbasi2018gamma,belz2020observations}.

The  \textbf{ High-speed video camera} is a monochrome Phantom V2012 operating at 40,000 frames per second with a time interval between frames of 25.00 $\mu$s and an exposure time of 23.84 $\mu$s (at the end of each frame the camera is blind for 1.14 $\mu$s due to data transfer). Each frame of the video is time-stamped by utilizing a GPS antenna and has a resolution of 1280 $\times$ 448 pixels. The camera is sensitive to the visible and  near-infrared spectra (400 nm - 1000 nm). The camera was installed five kilometers to the east border of the TASD inside a building adjacent to the interferometer (INTF) and the electric-field fast antenna (FA). The 20-mm focal length lens allowed a vertical viewing angle of 35 degrees, and a horizontal angle of 84 degrees covering almost all of the TASD detectors. The camera's position and settings were optimized to observe downward TGF sources that are approximately 30 km from the camera and up to 3 km above ground level. Each video had a recording length of 1.1 seconds and was automatically triggered by changes in luminosity. Data from the camera is stored on a computer at the site and analyzed offline. The elevation angles from the camera were used to calculate the source height of the TGF sources.

  The \textbf{ Lightning Mapping Array (LMA)}, developed by the Langmuir Laboratory group at New Mexico Tech~\cite{rison1999gps,thomas2004accuracy},  has been running at the TASD detector since 2013.  The LMA detects the lightning sources emitting impulsive signals between 60MHz -66MHz. LMA-detected sources are analyzed using the time-of-arrival technique of the impulse signal time for multiple triggered LMA stations on the ground. This technique provides us with detailed 3D images of peak VHF radiation events above threshold in 80 $\mu$s time intervals. The LMA detects VHF peak with a time accuracy of 35-ns root mean square over a wide  ($>$ 70 dB) dynamic range, from less than 10 mW to greater than 100 kW peak source power~\cite{thomas2004accuracy}.   
  
 The \textbf{INTerFerometer (INTF)}  and the \textbf{Fast electric field change Antenna (FA)} at the TASD site have been running since 2018.  The INTF records broadband (20 - 80 MHz) waveforms at 180 MHz from three flat-plate receiving antennas.  The three antennas were positioned in a triangular baseline of 106–121 m. Such a baseline was used to maximize the angular resolution over the TASD detector. The INTF is processed offline to determine the two-dimensional azimuth and elevation arrival directions of the VHF radiation with sub-microsecond resolution~\cite{stock2014continuous}.  The FA, on the other hand, records the electric field changes of lightning discharges. The FA provides high-resolution 180 MHz measurements of the LF/ELF discharge sferics~\cite{Liu2019,belz2020observations}. The FA uses a downward-looking flat plate sensor. The FA data is stored locally and processed offline similar to the INTF data.

\section*{Observations and Analysis}

After the installation of the high-speed video camera on 12 August 2021, we recorded several lightning flashes over the TASD before detecting the first optical observation of of lightning discharges associated with downward-directed TGF events on 11 September 2021. On that day nine flashes occurred over the TASD detector. They were all cloud-to-ground flashes with negative polarity.
Six of these nine flashes produced TGFs, all of them occurring during a time interval of only 51 minutes. The three non-producing TGF flashes had return stroke peak currents of less than 26 kA as reported by the National Lightning Detection Network (NLDN). The six gamma-ray initiating flashes, on the other hand, were produced by flashes with peak currents of 51, 67, 53, 154, 134, and 223 kA consecutively as reported by the NLDN. In all the TGF-producing flashes, the TGFs were associated with the downward leader propagation before the first return stroke. Four of the TGF-producing flashes had multiple return strokes. All subsequent strokes had different ground contact points and the first return stroke was the most intense one.

While the TASD has observed about 27 TGFs between 2008-2020, this weakly convective, hail-producing storm, has been found to be an uncommon observation of a storm by the TASD. At first, all the TGFs observed on that day were  produced by cloud-to-ground flashes with return stroke peak currents that ranged in magnitude from 51 kA all the way to 223 kA. In addition, the maximum energy deposit on one of the surface detectors had reached energies of up to 33,913 Vertical Equivalent Muons (VEM) (74 GeV). Also, the duration of the observed TGF bursts reached up to 719 $\mu$s. Note that  previously detected TGFs by the TASD detector typically were produced by flashes with an average peak current of  52 kA and maximum peak current of 139 kA, a deposited maximum energy on a single surface detector of no more than 997 VEM (2 GeV) and a duration of less than 551~$\mu$s.  

Most importantly, while the average rate of TGF observations by the TASD detector is about two events per year, this storm resulted in six TGF observations within one hour (22\% of all TGF observed in the past 13 years). This makes it the highest rate of TGF observations  in both one thunderstorm and  in all thunderstorm seasons observed by the TASD detector in the southwestern desert of Utah.   

In this work, we focus on and make a detailed analysis of the fourth flash during which a TGF is observed at 17:11:12 UTC by the TASD, the high-speed video camera, INTF, and FA. In the following, we refer to this flash as TGF-4. A more detailed analysis of this special storm, and the other five TGF-producing flashes, is now underway and will be reported on in a future publication. We chose the TGF-4 event because it was the clearest and the most straightforward to analyze. This TGF resulted in a bursts of three gamma-ray triggers reported by the TASD detector. We will refer to these triggers as trigger A, B, and C. The gamma-ray footprint and LMA source locations for this TGF are shown in Figure~S2 and Figure~S3 in the supporting information. As shown in Figure~S2, the maximum size of the TGF footprint on the ground, as observed by the TASD, is approximately  6 km in diameter.  The maximum energy deposited in a TASD detector was 1.8 GeV. The TGF burst, which occurred during the propagation of a fast leader, lasted for 719 microseconds and was followed by a high peak current return stroke of -154~kA as reported by the NLDN.

\begin{figure*}[]

\includegraphics[width=1.1\textwidth]{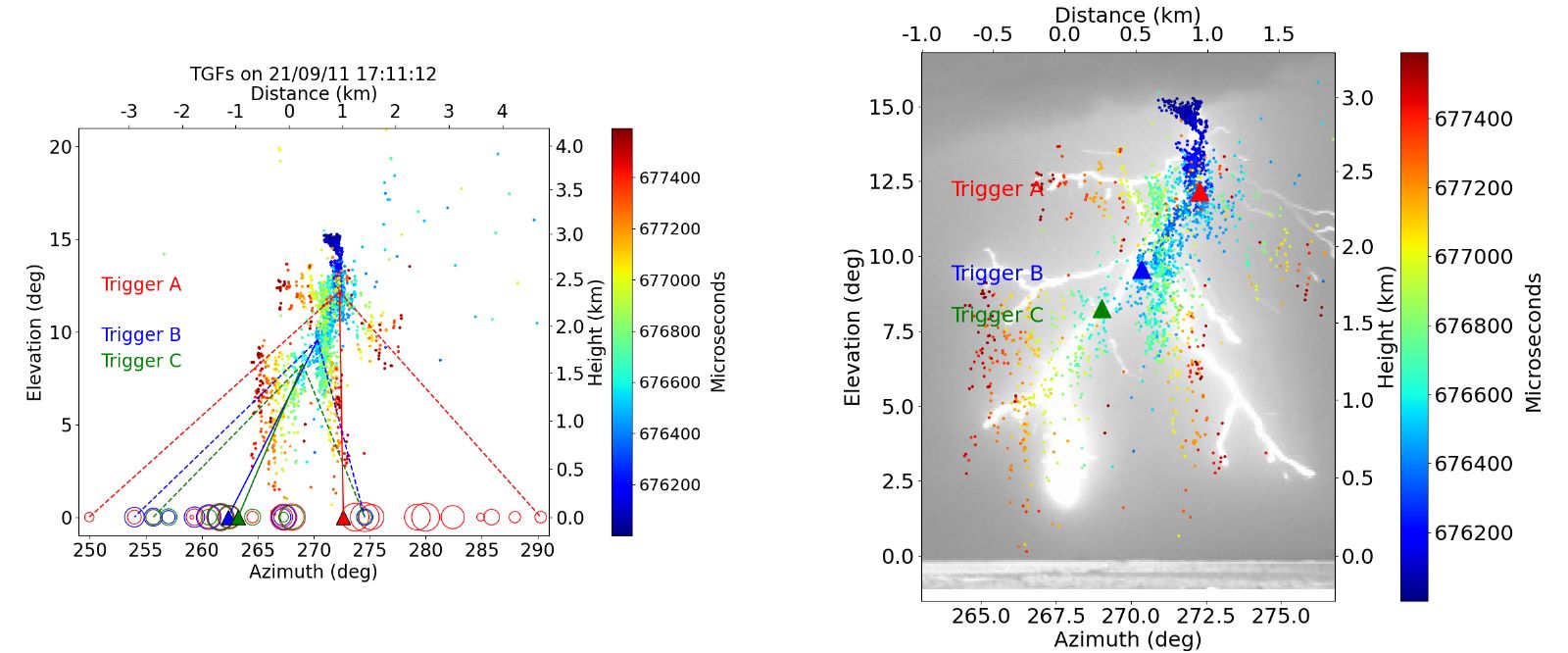}

\caption{The left-hand plot shows the elevation vs. azimuth plots of the INTF observations. The color represents the timing (blue is earlier and red is later in the flash). The red, green, and blue dashed lines point from the source of the TGF for trigger A, B, C and consecutively  to the TASD detectors at the edges of the footprint observation on the ground.  The circles on the ground refer to the TASD detectors triggered by the TGF observation. The filled circles show the central position of the footprint weighted by the energy deposited in the TASD. The color corresponds to each trigger using the same color code as the dashed lines.  The size and color of each circle are proportional to the energy deposited in the scintillator detector logarithmically.  The right-hand plot shows the elevation vs. azimuth for the whole flash in a frame of the camera in addition to the INTF point sources using the same color scale as the left-hand plot. The filled triangles indicate to the source height for each trigger obtained from the iteration procedure.}
\label{fig:camera1}
\end{figure*}

 The TGF-4-burst  A, B, and C trigger sources were produced over the eastward part of the TASD detector with a 10.9 km distance from the high-speed video camera, INTF, and FA location.  Trigger A, B, and C source heights were obtained using two independent analyses and were found to be consistent within uncertainties. In the first method, the heights were found using the camera pointing direction and the distance given by the LMA, resulting in heights of 2.5~km, 1.9~km, and 1.3~km above ground level for triggers A, B, and C respectively. In the second method, the heights were found using the INTF elevation and azimuthal direction and the distance given by the LMA, utilizing the iteration procedure used in Belz et al., 2020, resulting in heights of 2.4~km, 1.9~km, and 1.6~km above ground level. Moreover, the propagational two-dimensional speeds of the leader at triggers A, B, and C were found to be  7.2$\times 10^{6}$ m/s,  2.5 $\times 10^{6}$ m/s, and 3.0 $\times 10^{6}$ m/s .
The speed values reported above are higher than the average  stepped leader speeds~\cite{CAMPOS2014285}. It is also found to be higher than  average upward development speeds of intra-cloud discharges during the production of  TGFs~\cite{shao2010,cummer2015}. Figure~\ref{fig:camera1} shows the lightning flash and the height, elevation, and azimuth of the trigger A, B, and C sources as observed by the high-speed video camera, the INTF, and the TASD detectors.

The fluence of this flash,  determined using a GEANT4 simulation, was estimated to be $3\times10^{15}$. The GEANT4 simulation used in this work is described in detail in~\citep{abbasi2018gamma}, where electrons above 100 keV are generated from a point-like source, according to a RREA spectrum, propagated through the atmosphere, and the TASD detector. Based on our observation of the shower footprint, source position, and leader altitudes, particles are assumed to be forward beamed within a cone of half-angle of $20^{\circ}$. %This is the highest fluence at the source, to this date, determined from GEANT4 simulation, for downward-TGF observations detected at the TA site.

\begin{figure*}[]
\begin{center}
\includegraphics[width=0.5\textwidth]{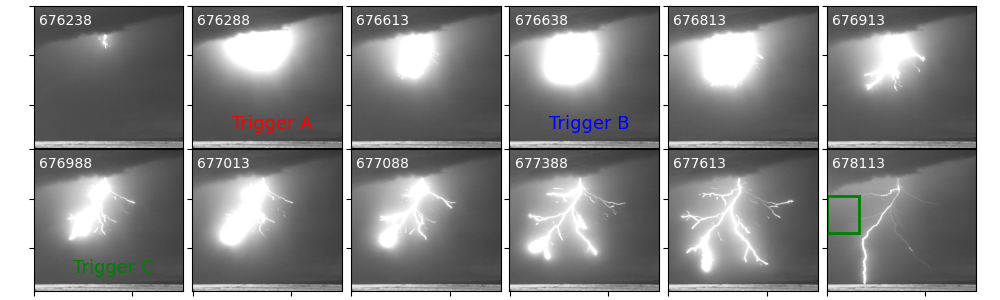}
\caption{ High-speed video selected frames of the TGF producing flash from the time it breaks below the cloud until the return stroke.  Note that the time difference between these images does not correspond to the time resolution of the camera. These images are 12 images  selected to cover the view of the whole flash (more images from the camera are shown in Figure S5 in the supporting information). The green square over the last image indicates the area monitored to extract the luminosity variation. The timing in microseconds from  17:11:12 UTC is displayed at the left corner of each image.}
\label{fig:camera2}
\end{center}
\end{figure*}

The optical emission, in this work, was analyzed using the high-speed video camera. Figure~\ref{fig:camera2} shows the progression of the leader of this flash in multiple selected frames as observed by the high-speed video camera. Some of the selected frames display a light saturation problem. This problem is due to the fact the camera's settings were conservatively optimized for suspected multiple scenarios including a very faint signal if the sources of the TGFs were inside the cloud and would suffer from light scattering. The initial camera settings would have, in principle, allowed us to detect  possible faint signals due to scattering or absorption for up to 30 kilometers away. However, this TGF storm was the most energetic TGF-producing thunderstorm we have ever observed. The fact that the TGF signals were associated with a well-developed, very intense leader which was below the cloud base
contributed to the high luminosity. The intense luminosity, high leader speed, and large return-stroke current suggest that the leader had a very high charge density. In two cases (trigger B and C) the leader was almost halfway to the ground at the time of the gamma-ray emissions, as shown in Figure~\ref{fig:camera1}.  

\begin{figure}[]
\includegraphics[width=0.5\textwidth]{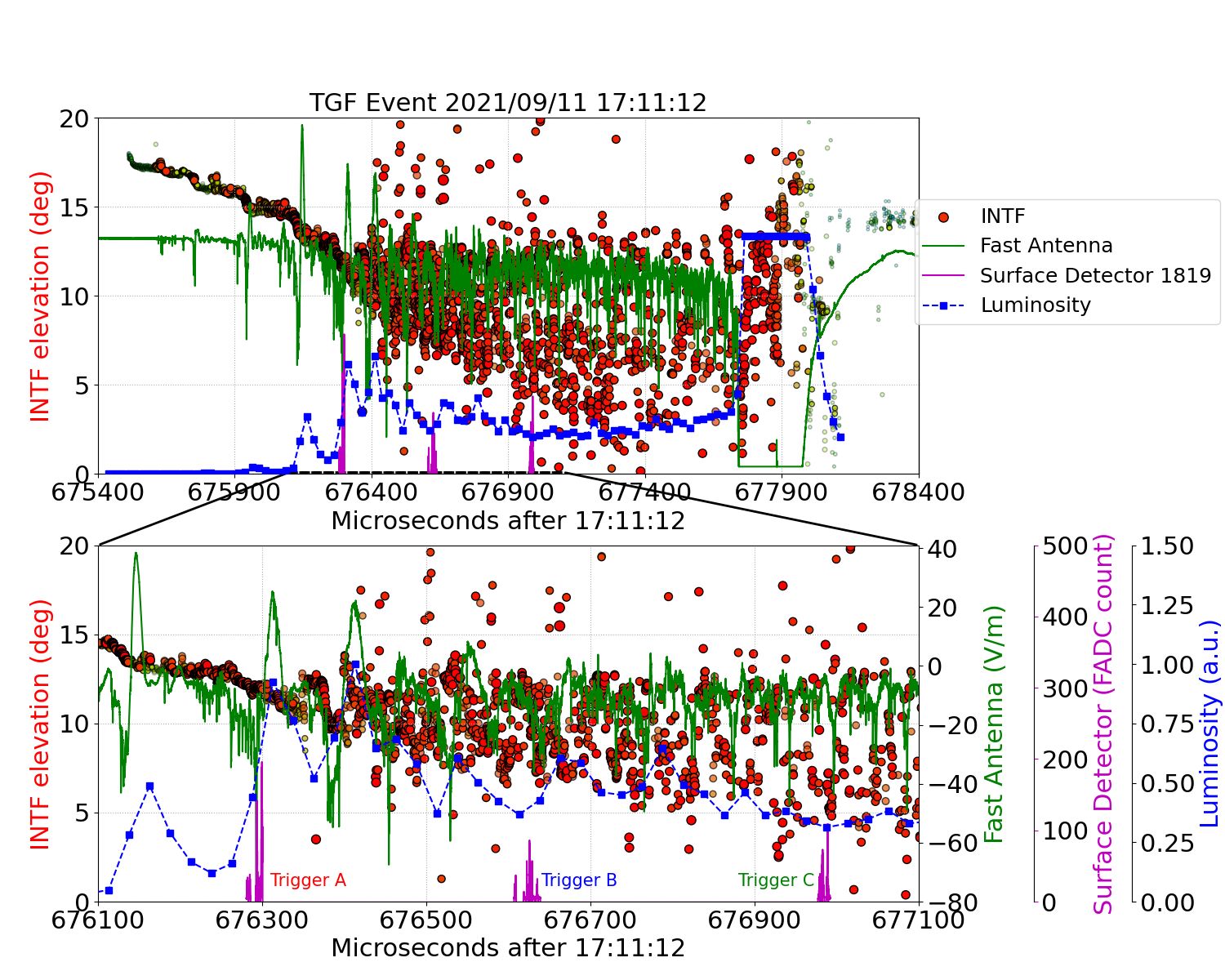}
\caption{ This figure shows the TASD waveforms for one of the SDs (1819) in magenta, the average luminosity in dark blue, the electric-field waveform in green, the INTF elevation in red circles (size and color are proportional to the power of the radio signal). Top: The flash observed from initiation until the first return stroke within 3 ms duration. Bottom: a zoomed in version of the top plot within 1 ms. added the optical emission of the return stroke in the top figure.}
\label{fig:intf}
\end{figure}

Figure~\ref{fig:intf} combines the TASD, high-speed video camera luminosity, INTF, and FA observations. The TGF timing relative to the INTF, electric field change pulses, and visible light emission was calculated using the time matching analysis between the TASD, the INTF, and the LMA as carefully described in detail in~\citep{belz2020observations}.  Two different zooms of this flash are displayed. The top panel of Figure~\ref{fig:intf} shows three milliseconds of the detectors’ observations, while the bottom panel shows a one-millisecond zoom. The TASD, INTF, and FA data displayed in this plot are both triggered and analyzed similarly to previous TGFs we  reported on in~\citep{belz2020observations}. Note that the first electric field breakdown pulse (at 675754 $\mu$s in Figure 3-top) occurs well before (515 $\mu$s) the first TGF burst (at 676269 $\mu$s). The first breakdown pulse, according to the INTF, appears to occur at an approximate height of 3400~m. However, at the time of the first TGF burst, the leader tip is already well below the cloud base (second frame in Figure~\ref{fig:camera2}).

The luminosity, shown in dark blue, is measured by averaging the pixel values of an area adjacent to the lightning flash channel (diffused, not direct luminosity) in each video image. This area is indicated by a green rectangle shown in the last frame presented in Figure~\ref{fig:camera2}.  The motivation behind this method is to avoid misleading peaks in the luminosity (possible plateaus) that  could have been encountered due to the saturation of pixels in several snapshots in Figure~\ref{fig:camera2}. Note that we did compare both the average luminosity, including the adjacent pixels to the flash vs. the pixels including the flash itself, and found that both intensity curves are consistent with each other (shown in Figure~S4 in the supporting information).

%%%%%%%%%%%%%%%%%%%%%%%%%%%%%%%%%%%%%%
%\input{analysis}

\section*{Discussion}

This work presents the first simultaneous detection of a downward TGF together with the observation of the associated cloud-to-ground lightning flash by a high-speed video camera.  The camera allowed us to check the development stage and the luminosity of the lightning leader during the occurrence of an extremely energetic downward-direct terrestrial gamma-ray flash. The detected TGF presented unique features in terms of energy deposit, and duration, and was associated with an unusually fast stepped leader that produced a very high peak current return stroke. Unlike most of the other TASD observed events~\cite{abbasi2018gamma, belz2020observations}, this extremely energetic downward-direct terrestrial gamma-ray is not solely related to the early leader stage. The energetic observed gamma-ray triggers were produced as the leader clearly propagated below the cloud base. The third TGF burst happened when the stepped leader was approximately halfway to the ground. Observing a TGF during the stepping leader further unlocks the circumstances under which TGFs are produced. Such observation may shed light on the lightning-stepping process itself. 

X-rays have also been detected during the propagation of energetic stepped and dart leaders to ground in past studies (e. g.~\cite{Moore2001,Howard2019,Saba2019,Urbani2021}). The typical downward-directed TGF observations by the TASD are from the initial leader stage of lightning flashes 3-5 km above ground level. Other experimental works have also reported TGF observations  following return strokes of -CG discharges~\cite{dwyer2012b,https://doi.org/10.1002/2013JA019112,Tran2015}. X-rays, on the other hand, appeared to be detected during the propagation of energetic stepped and dart leaders a few hundred meters above the ground. X-rays have been observed to have a  softer energy spectrum than gamma-rays which are not more than~250 keV~\cite{dwyer2004a}. TGFs and X-rays are the two most common energetic radiation during a lightning process. Until now, it is not clear whether TGFs and X-ray emissions detected at ground level are related. That being said, it is important to reiterate, and as discussed in detail in our previous publications~\cite{abbasi2018gamma,belz2020observations}, that the TASD detects multi-MeV gamma-radiation and is blind to X-radiation.

%TGFs and X-rays are the two most common energetic radiation during a lightning process. The typical downward-directed TGF observations by the TASD are from the initial leader stage of lightning flashes 3-5 km above ground level. Other experimental work have also reported TGF observations  in the early period of a return stroke~\cite{Tran2015,Hare2016,Enoto2017}. X-rays, on the other hand, appeared to be detected during the propagation of energetic stepped and dart leaders a few hundred meters above the ground~\cite{Moore2001,Howard2019,Saba2019,Urbani2021}). X-rays have been observed to have a  softer energy spectrum than gamma-ray which are not more than~250 keV~\cite{dwyer2004a}. Until now, it is not clear whether TGFs and X-ray emissions detected at ground level are related. That being said, it is important to reiterate, and as discussed in detail in our previous publications~\cite{abbasi2018gamma,belz2020observations}, the TASD is detecting multi-MeV gamma-radiation and is blind to x-radiation.}

The measured luminosity of the optical emissions observed by the high-speed video camera were found to start to increase within 25 microseconds from the onset of the TGF trigger signal and peak some 25-75 microseconds after the TGF trigger ceases.  This may be coincidental, or it could be part of a pattern.  We intend to study this further by examining the other TGF observations we made and continue to make at the TASD site, and will report on our findings in future publications.  It is interesting to note that~\citep{belz2020observations}  showed that TGF triggers were associated with an electric field pulse (initial breakdown pulse) which was not necessarily the largest one, and that, not all initial breakdown pulses were associated with TGF triggers. Similarly, in Figure~\ref{fig:intf}, we observed that some luminosity peaks occurred after the TGF bursts. They were not necessarily the largest ones and not all the luminosity peaks were associated with TGF triggers.

ASIM collaboration reported on TGF detections with associated optical pulses~\cite{asim10m,skie22}. They found that the majority are observed to start after a weak increase in the optical emission in the 337 nm and 777.4 nm photometers, and before, or at the onset of, the main $\sim$2 ms long optical pulse~\cite{asim10m}. In our data, we observe the first TGF trigger roughly $200~\mu s$ after the onset of the main optical pulse, as can be seen in Figure~\ref{fig:intf}. However, we note that the TGF occurs near the start of the main optical pulse, which lasts for about 2~ms, and that the optical luminosity increases and peaks after the first TGF trigger.  
Future analysis of the distribution of the relative time difference between TGFs and the main optical pulses, observations of  the UV emissions,  and simulation work are planned to better understand this issue.

\section*{Summary and Future Outlook}

On September 11 of 2021, we  observed the highest TGF-rate-producing thunderstorm by the TASD. This storm was responsible for a significant fraction of all ground TGFs detected over the TASD in the past ten years. This work presents the first simultaneous detection of one of those downward-directed TGFs together with the observation of the associated cloud-to-ground lightning flash by a high-speed camera, an INTF, and a FA. In this paper, we investigate the optical emission of a lightning flash associated with  an energetic downward TGF. The TASD in addition to the suit of lightning instruments including a high-speed video camera allowed us to understand in detail the height, speed, footprint, energy deposit, and stage of lightning in the flash that is associated with gamma-ray bursts observed by the TASD detector.

Based on the simultaneous use of the TASD, lightning instruments with a high-speed camera recording 40,000 images per second, we observed that:
\begin{enumerate}
\item	The energetic detected TGF burst occurred during the propagation of a fast and bright downward negative leader which resulted in a high peak current return stroke of -154 kA.
\item	The leader was propagating below the cloud base during the TGF burst. The last trigger occurred when it was almost halfway to the ground.
\item The TGF  presented unique features in terms of energy deposit and duration.
\item The TGF was produced in an uncommon storm at the TASD site. This single storm was responsible for the detection of a significant fraction of all ground TGFs detected over ten years.

\end{enumerate}
 To understand the physics behind the initiation and propagation of TGFs, and to further compare the optical signature correlation between downward and upward-directed TGF emissions (Ostgaard et al., 2019; Heumesser et al., 2021; Lindanger et al., 2022; Bjorge-Engeland et al., 2022) we have installed photometers at the TASD site. These photometers share the same field of view as the high-speed video camera and will report, with much higher timing resolution, about the optical emissions from atmospheric electrical discharge processes in three different wavelengths: at 337.0~nm (associated with the second positive system of the nitrogen molecule ($2PN_2$)), 391.2~nm (associated with $1NN2^+$ emission correspond to the first negative system of the nitrogen molecule $N2^+$), and 777.4~nm (associated with the atomic oxygen ($OI$)). This will hopefully allow us to build further conclusions that support the model responsible for the TGF initiation and verify if downward-directed TGFs are a variant of the same phenomenon that causes upward-directed TGFs.

\section*{Acknowledgements}

 Operation and analyses of this study have been supported by NSF grants AGS-2112709, AGS-1844306, AGS-1613260, AGS-1205727,  and AGS-1720600. The Telescope Array experiment is supported by the Japan Society for
the Promotion of Science(JSPS) through
Grants-in-Aid
for Priority Area
%"Highest Energy Cosmic Rays"
431,
for Specially Promoted Research
%``Extreme Phenomena in the Universe Explored by Highest Energy Cosmic Rays''
%Grant Number
JP21000002,
%Grant-in-Aid
for Scientific  Research (S)
%"Quest for the unified picture of the explosion mechanism of supernovae and the central engine of gamma-ray bursts"
%Grant Number
JP19104006,
%Grant-in-Aid
for Specially Promoted Research
%"Extended Telescope Array Experiment - Nearby Extreme Universe Elucidated by Highest-energy Cosmic Rays"
%Grant Number
JP15H05693,
%Grant-in-Aid
for Scientific  Research (S)
%"Study of the ultra high energy cosmic ray source evolution by detailed measurement of cosmic rays in the wide energy range"
%Grant Number
JP15H05741, for Science Research (A) JP18H03705,
%Grant-in-Aid
for Young Scientists (A)
%"hoge hoge"
%Grant Number
JPH26707011,
and for Fostering Joint International Research (B)
%"Search for Ultra-High Energy Cosmic Ray origin using the extended Telescope Array experiment"
%Grant Number
JP19KK0074,
by the joint research program of the Institute for Cosmic Ray Research (ICRR), The University of Tokyo;
by the U.S. National Science
Foundation awards PHY-0601915,
PHY-1404495, PHY-1404502, and PHY-1607727;
by the National Research Foundation of Korea
% \linebreak
(2016R1A2B4014967, 2016R1A5A1013277, 2017K1A4A3015188, 2017R1A2A1A05071429) ;
%\linebreak
by the Russian Academy of
Sciences, RFBR grant 20-02-00625a (INR), IISN project No. 4.4502.13, and Belgian Science Policy under IUAP VII/37 (ULB). The foundations of Dr. Ezekiel R. and Edna Wattis Dumke, Willard L. Eccles, and George S. and Dolores Dor\'e Eccles all helped with generous donations. The State of Utah supported the project through its Economic Development Board, and the University of Utah through the Office of the Vice President for Research. The experimental site became available through the cooperation of the Utah School and Institutional Trust Lands Administration (SITLA), U.S. Bureau of Land Management (BLM), and the U.S. Air Force. We appreciate the assistance of the State of Utah and Fillmore offices of the BLM in crafting the Plan of Development for the site.  Patrick Shea assisted the collaboration with valuable advice  on a variety of topics. The people and the officials of Millard County, Utah have been a source of steadfast and warm support for our work which we greatly appreciate. We are indebted to the Millard County Road Department for their efforts to maintain and clear the roads which get us to our sites. We gratefully acknowledge the contribution from the technical staffs of our home institutions. An allocation of computer time from the Center for High Performance Computing at the University of Utah is gratefully acknowledged. We thank Ryan Said and W. A. Brooks of Vaisala Inc. for providing quality NLDN data lightning discharges over and around the TASD under their academic research use policy.

%%%%%%%%%%%%%%%%%%%%%%%%%%%%%%%%%%%%%%%%%%%%%%%%%%%%%%%%%%%%%%%%%%%%%%%%%%%%%%%%
\section*{Open Research}

The data used in this paper are available through this link: http://doi.org/10.17605/OSF.IO/WH4UP. They are uploaded in the following directories: The Fast Antenna (FA), the INTerFerometer (INTF), the Lightning Mapping Array (LMA), the high-speed video camera (Optical data), and the Telescope Array Surface Detector (TASD). To be able to look at the optical data, a PCC 3.6 Phantom software is needed. The package was uploaded in the same directory. 

%%%%%%%%%%%%%%%%%%%%%%%%%%%%%%%%%%%%%%%%%%%%%%%%%%%%%%%%%%%%%%%%%%%%%%%%%%%%%%%%
\bibliography{paper1}{}
\bibliographystyle{ieeetr}

%\bibliography{paper1}{}
%\bibliographystyle{ieeetr}

%%%%%%%%%%%%%%%%%%%%%%%%%%%%%%%%%%%%%%%%%%%%%%%%%%%%%%%%%%%%%%%%%%%%%%%%%%%%%%%%
\clearpage

\section{ Supporting Information}
\clearpage

%%%%%%%

\begin{figure}[]
\begin{center}
\includegraphics[width=0.4\textwidth]{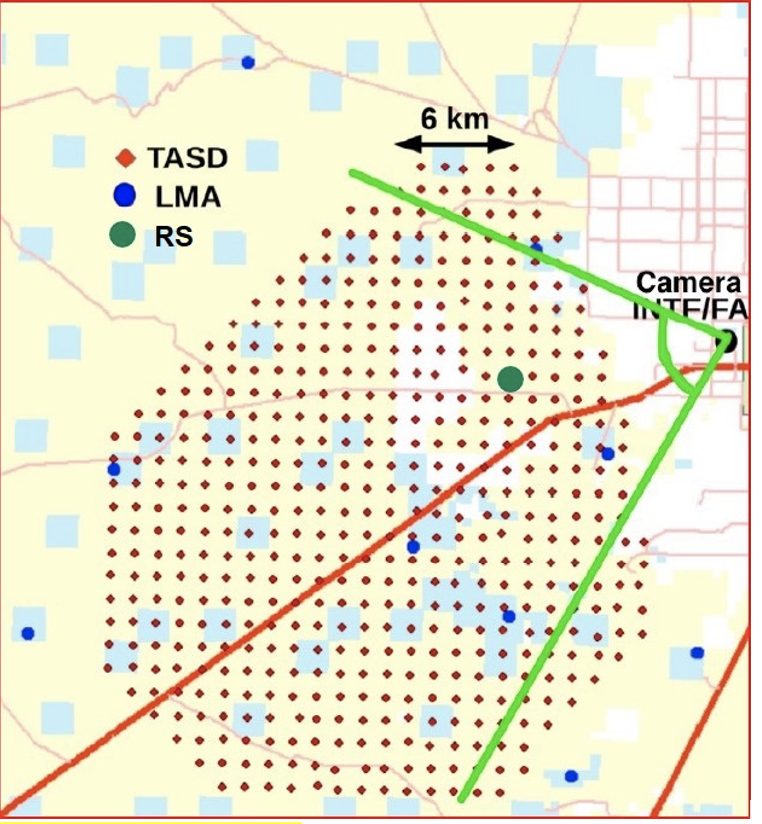}
\caption{ The layout of the instrumentation and the position of the return stroke (RS) associated with the TGF bursts shown in dark green circle. The 507 stations of the Telescope Array Surface Detector (TASD) are shown as red diamonds, and the eleven Lightning Mapping Array (LMA) stations as blue circles. The high-speed video camera, the interferometer (INTF) and the Fast Antenna (FA) are located five kilometers to the eastern most edge of the TASD. The field of view of the high-speed video camera is shown by the
bright green lines with an opening angle of 84 degrees.}
\label{fig:layout}
\end{center}
\end{figure}

\begin{figure}[]
\begin{center}
\includegraphics[width=0.4\textwidth, angle=0]{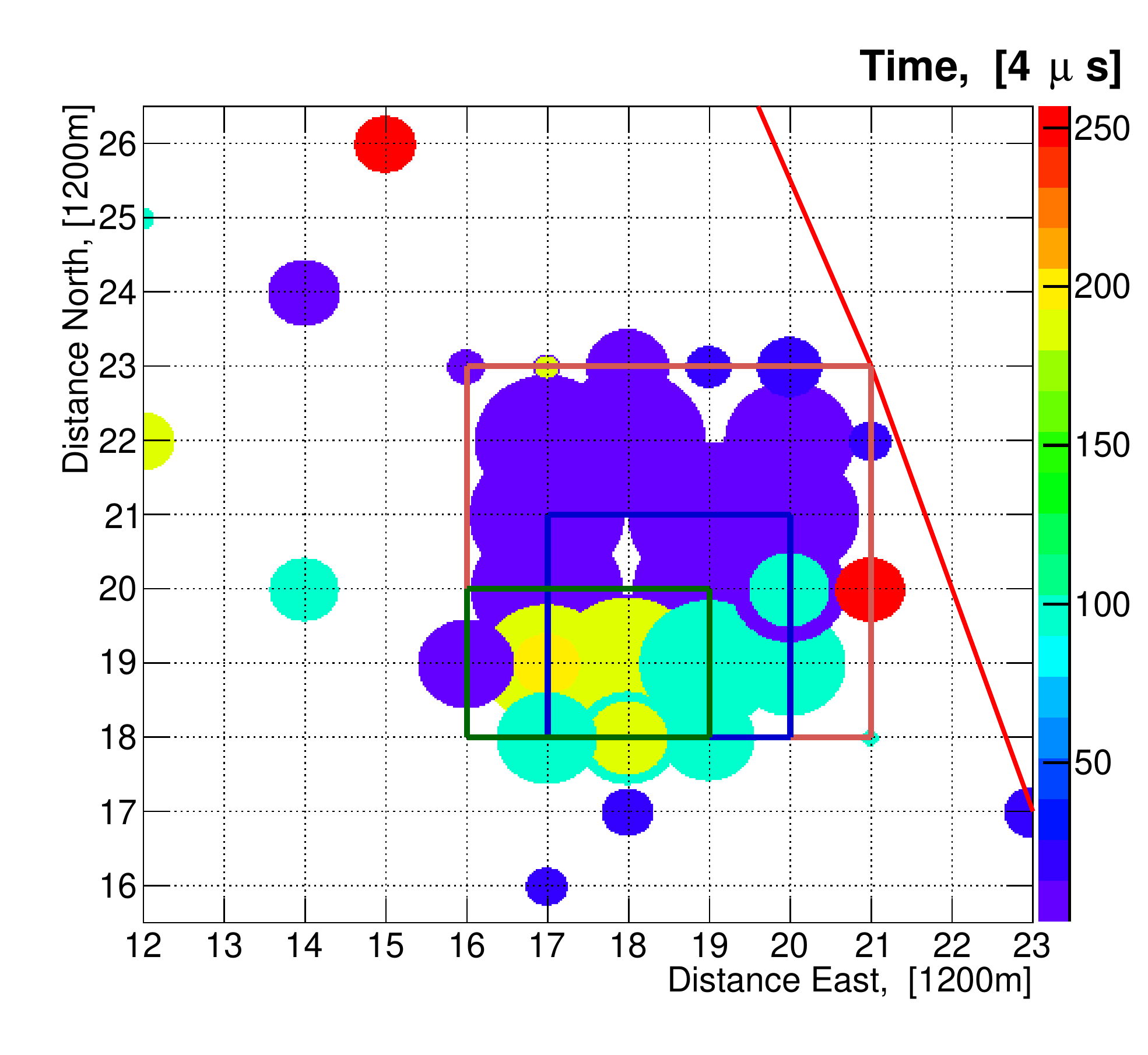}
\caption{ The TASD footprint including trigger A, trigger B, and trigger C bounded by the red, green, and yellow squares consecutively. The grid spacing of the surface scintillators is 1200~m. The area of each circle is proportional to the logarithm of the energy deposit, and the color indicates relative timing in 4~$\mu$s steps. The red line denotes the eastern border of the TASD array. }
\label{fig:display}
\end{center}
\end{figure}

\begin{figure}[]
\begin{center}
\includegraphics[width=0.5\textwidth]{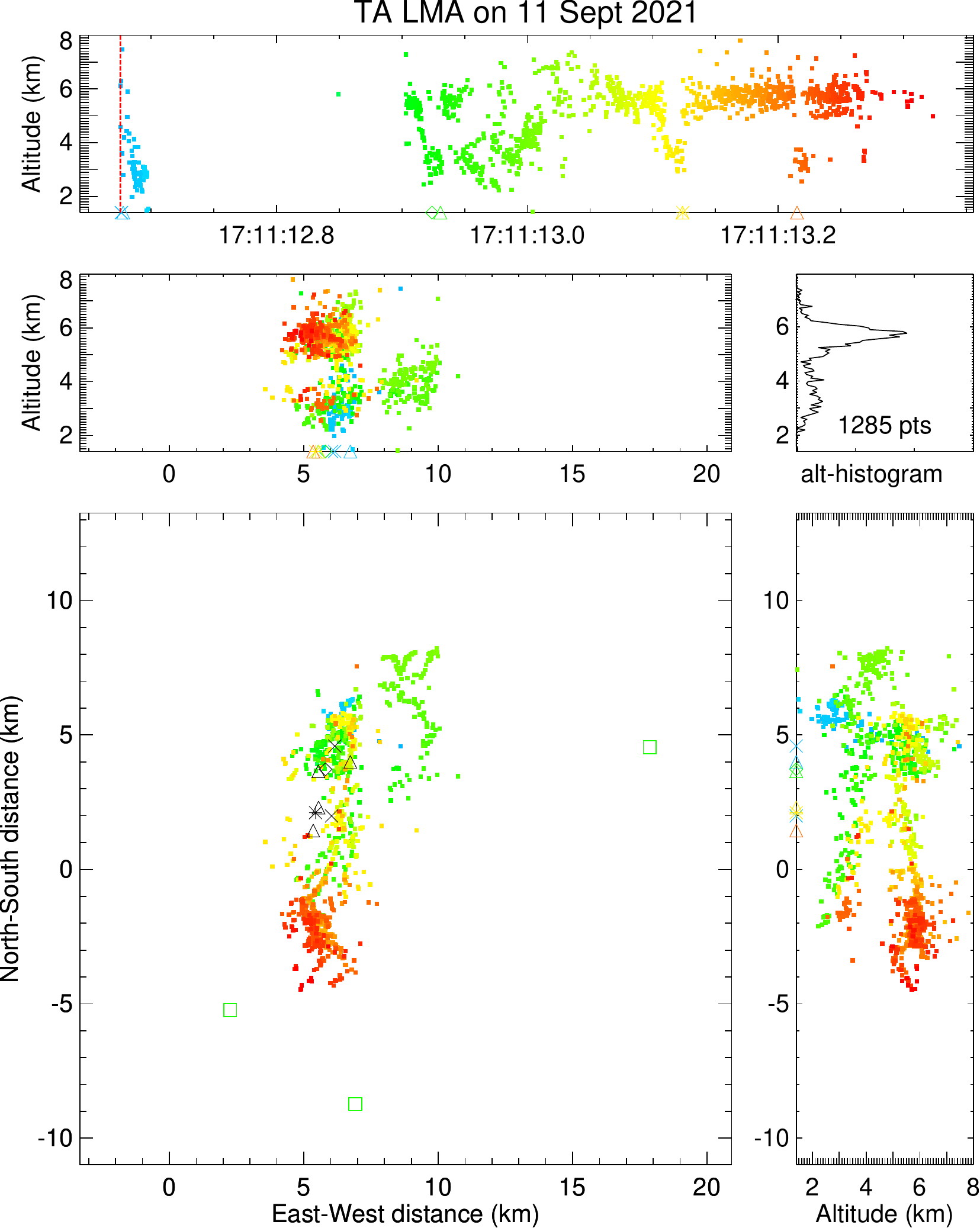}
\caption{  LMA source locations of flash 17:11:12 on 11 Sept. 2021 indicated by filled square symbols
color-coded by time.  The vertical dashed red line shows the time of the TGF bursts.  Ground
is at 1.4 km altitude, corresponding to the lower axis of the height-time and vertical
projection panels. National Lightning Detection Network events are shown as triangles for
-CG strokes, a diamond for a -IC stroke, an asterisk for a +IC stroke, and Xs for +IC
strokes which were mis-classified by NLDN as +CG strokes.  The green squares show the
locations of three of the eleven LMA stations.}
\label{fig:lma}
\end{center}
\end{figure}

\begin{figure}[]
\begin{center}
\includegraphics[width=0.5\textwidth]{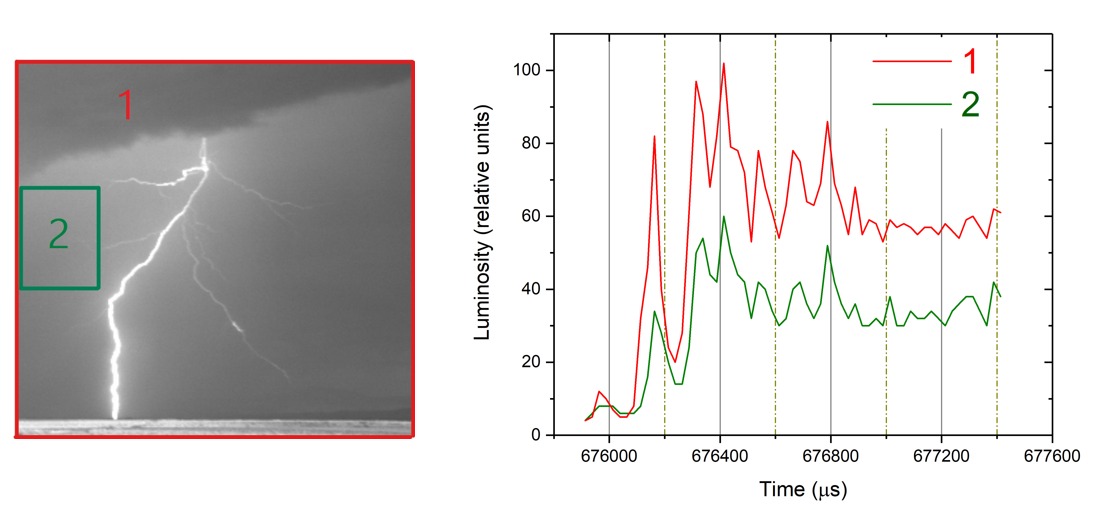}
\caption{ The left panel shows an image from the high-speed video camera, and highlights two regions. The red square (region 1) includes all pixels of the image and the green square (region 2) only pixels that were not saturated by the lightning flash. The right panel shows the luminosity vs. time for both regions. The luminosity is calculated by summing over all the pixel intensity values and then dividing the sum by the total number of pixels, inside the corresponding square.}
\label{fig:lma}
\end{center}
\end{figure}

%\begin{figure*}[]
%%\includegraphics[width=0.35\textwidth]{Camera_overall}
%\includegraphics[width=0.5\textwidth]{Figure3_zoom1.png}
%\includegraphics[width=0.5\textwidth]{Figure3_zoom2.png}
%\includegraphics[width=0.5\textwidth]{Figure3_zoom3.png}

%\caption{ This figure shows the TASD waveforms in hot pink, the average luminosity  vs. time  in dark blue,  the fast antenna waveform in green, the INTF elevation vs. time in red and black circles (Only the black circles are used in the velocity calculation). Top: The flash observed zoomed in on trigger A. Middle: The flash observed zoomed in on trigger B. Bottom: The flash observed zoomed in on trigger C. The TGF propagation speed is indicated by the solid black line and associated value.}
%\end{figure*}

\begin{figure}[]
\begin{center}
\includegraphics[width=0.7\textwidth]{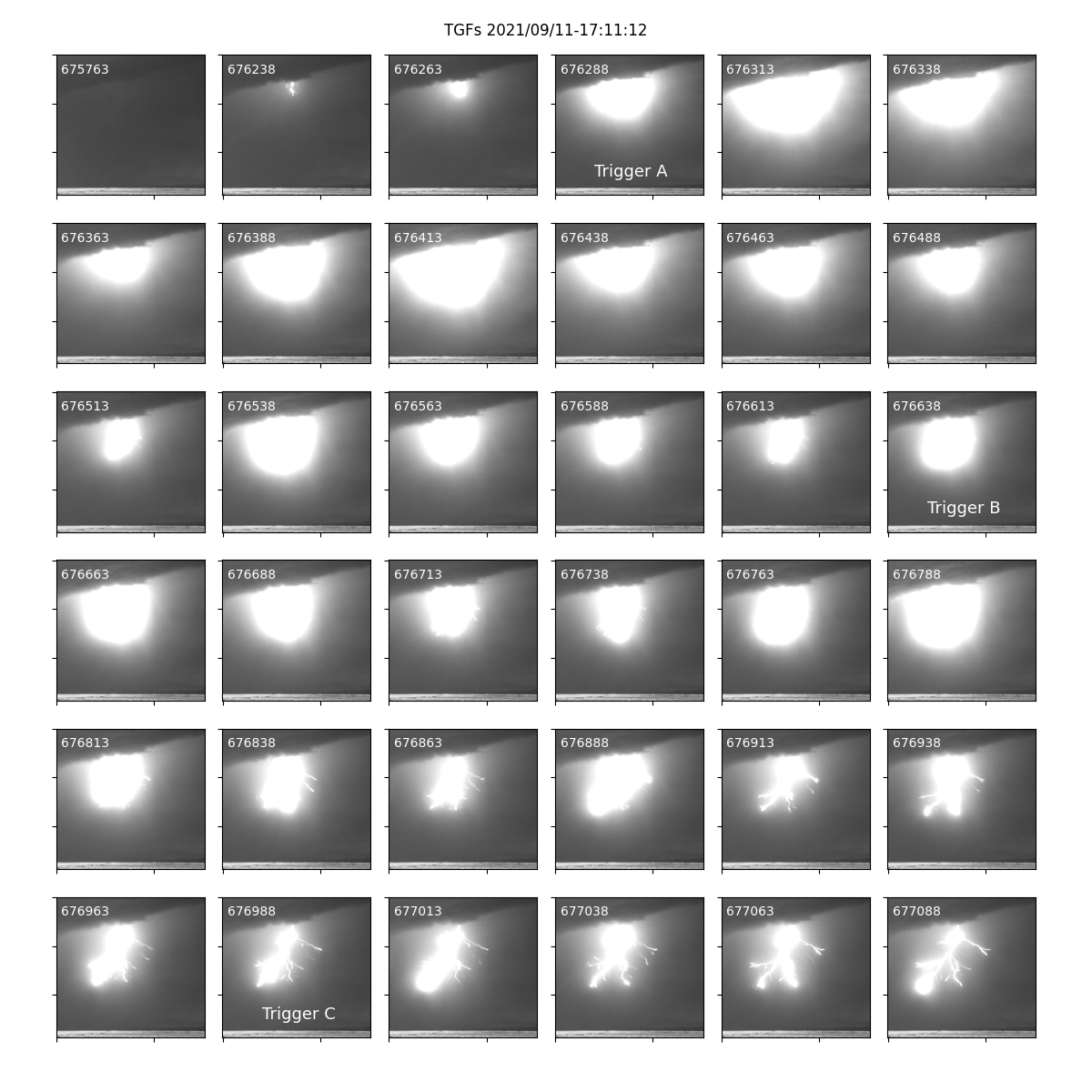}
\caption{ Selected high-speed video frames of the TGF producing flash from the moment of the first breakdown pulse until the return stroke occurence. }
\label{fig:lma}
\end{center}
\end{figure}

%%%%%%

%%%%%%%%%%%%%%%%%%%%%%%%%%%%%%%%%%%%%%%%%%%%%%%%%%%%%%%%%%%%%%%%%%%%%%%%%%%%%%%%
%\input{support}
\end{document}